\documentstyle[12pt]{article}

\def\hybrid{\topmargin -20pt  \oddsidemargin 0pt
      \headheight 0pt   \headsep 0pt
      \textwidth 6.25in 
      \textheight 9.5in 
      \marginparwidth .875in
      \parskip 5pt plus 1pt   \jot = 1.5ex}

\hybrid

\begin{document}
\def\mgz{{\cal M}^G_Z}
\def\mge{{\cal M}^G_E}
\def\mgeb{{\cal M}^G_{E_b}}
\def\mgzb{{\cal M}^G_{Z/B}}
\def\mtz{{\cal M}^T_Z}
\def\mte{{\cal M}^T_E}
\def\x{\times}
\def\ra{\rightarrow}
\def\beq{\begin{equation}}
\def\eeq{\end{equation}}
\def\beqa{\begin{eqnarray}}
\def\eeqa{\end{eqnarray}}

\sloppy
\newcommand{\be}{\begin{equation}}
\newcommand{\eq}{\end{equation}}
\newcommand{\ov}{\overline}
\newcommand{\tl}{\widetilde}
\newcommand{\un}{\underline}
\newcommand{\p}{\partial}
\newcommand{\la}{\langle}
\newcommand{\ds}{\displaystyle}
\newcommand{\nl}{\newline}
\newcommand{\th}{\theta} 
\parindent1em

\begin{titlepage}
\begin{center}

\vskip .7in  

{\bf ICMP lecture on Heterotic/F-theory duality}

\vskip .3in

Ron Donagi \noindent \footnote{email:  donagi@math.upenn.edu\\ 
Partially supported by NSF grant DMS 95-03249 and 
(while visiting ITP) by NSF grant
PHY94-07194.}  \\ \vskip 1.2cm

{\em  Department of Mathematics, 
University of Pennsylvania, Philadelphia, PA 
19104-6395}

\vskip .1in

\end{center}

\vskip .2in

\begin{quotation}\noindent

ABSTRACT 

The heterotic string compactified on an $n-1$-dimensional elliptically fibered
Calabi-Yau  $\pi_H:  Z  \ra  B$  is  conjectured  to  be  dual  to  $F$-theory
compactified on an $n$-dimensional Calabi-Yau $\pi_F:X \ra B$ fibered over the
same base with  elliptic  K3  fibers.  In  particular,  the  moduli of the two
theories  should be  isomorphic.  The cases most  relevant to the  physics are
$n=2,  3, 4$,  i.e.  the  compactification  is to  dimensions  $d=8,6$  or $4$
respectively.  Mathematically,  the richest picture seems to emerge for $n=3$,
where the moduli space involves an analytically integrable system whose fibers
admit rather different  descriptions in the two theories.  The purpose of this
talk is to review  some of what is known and what is not yet known  about this
conjectural isomorphism. Some of the underlying mathematics of principal 
bundles on elliptic fibrations is reviewed in the accompanying Taniguchi  talk.

\end{quotation}
\end{titlepage}
\vfill
\eject

\section{Introduction}

The heterotic string compactified on an $n-1$-dimensional elliptically fibered
Calabi-Yau  $\pi_H:  Z  \ra  B$  is  conjectured  to  be  dual  to  $F$-theory
compactified on an $n$-dimensional Calabi-Yau $\pi_F:X \ra B$ fibered over the
same base with  elliptic  K3  fibers.  In  particular,  the  moduli of the two
theories  should be  isomorphic.  The cases most  relevant to the  physics are
$n=2,  3, 4$,  i.e.  the  compactification  is to  dimensions  $d=8,6$  or $4$
respectively.  Mathematically,  the richest picture seems to emerge for $n=3$,
where the moduli space involves an analytically integrable system whose fibers
admit rather different  descriptions in the two theories.  The purpose of this
talk is to review  some of what is known and what is not yet known  about this
conjectural isomorphism.

Various aspects of heterotic  compactifications  on elliptic  Calabi-Yaus were
explored in  [\ref{FMW}],  [\ref{D}],  [\ref{BJPS}].  We emphasize the role of
cameral (and  spectral)  covers,  which were  introduced  in  [\ref{DD}],  and
applied to integrable  systems in [\ref{D2}]  and to the  heterotic  moduli in
[\ref{D}].  F-theory was created in [\ref{V}];  some of the ideas we need here
were  developed in  [\ref{MV}].  Results about the  comparison  of the ``base"
moduli in the two theories appear in [\ref{FMW}],  [\ref{BJPS}] as well as the
earlier  [\ref{BIKMSV}],  [\ref{AG}],  and  elsewhere.  The  comparison of the
``fiber" moduli appears in [\ref{CD}].

\section{Eight-dimensional compactifications}

In a  compactification  to any  dimenesion,  the heterotic  moduli include the
$(n-1)$-dimensional  elliptically  fibered Calabi-Yau $Z \ra B$ with a section
$\sigma :  B \ra Z$, together with a principal  $G$-bundle on $Z$, where $G=E8
\x E8$ or  $Spin(32)/{\bf  Z}_2$ (or one of their  subgroups),  and additional
fields  depending on the  dimension.  The F-theory  moduli  include the double
fibration  $X \ra B_F \ra B$, where  $B_F \ra B$ is a ${\bf  P}^1$-bundle  (we
will refer to $B,B_F$ respectively as the heterotic and F-theoretic bases) and
$X \ra B_F$ is an elliptic  fibration with section  $\sigma_F:B_F  \ra X$, and
additional fields depending on the dimension.

In an  eight-dimensional  compactification,  the base $B$ is a  point,  so the
heterotic CY is an elliptic curve $Z=E$ equipped with a pair of $E_8$ bundles,
while the F-theoretic CY is an elliptic K3 surface $X$.  It is well-known that
the moduli  spaces  ${\cal  M}_H, {\cal  M}_F$ for these two types of data are
both given by the Narain space [\ref{N}]
$$O(2,18;{\bf Z})\backslash O(2,18;{\bf R})/(O(2;{\bf R})\times 
O(18;{\bf R}))$$ 
In a sense, this is the basic form of the duality.  When compactified  further
to six or four  dimensions, the base of the integrable  system for each theory
will  parametrize  families  (one  or two  complex-dimensional)  of the  eight
dimensional data.  (The fiber data is new to these further compactificatiuons,
and will be handled differently.)

Because of the T-dualities, this  correspondence is not (algebro-)  geometric;
it involves the period map, which is only analytic.  A simple geometric recipe
for  recovering  $Z$ from $X$ for  $G=E_8  \x E_8$ was  nevertheless  given by
Morrison and Vafa [\ref{MV}]:  write $X$ in Weierstrass form
$$X:  \ \ y^2=x^3+xf+g$$ 
where $f,g$ are  sections  of ${\cal  O}(8),  {\cal  O}(12)$ on the base ${\bf
P}^1$.  In terms of a coordinate $t$ on ${\bf P}^1$, write
$$f=\sum_{i=1}^8 f_it^i, \ \ \ g=\sum_{j=0}^{12} f_jt^j.$$ 
Then the equation of the elliptic curve is given by the middle terms
$$Z:  \ \ y^2=x^3+xf_4+g_6.$$ 
Further, it is pointed out in [\ref{FMW}]  that the data of an $E_8$ bundle on
$Z$ is equivalent to embedding $Z$ in a rational  elliptic surface $dP_9$ with
a section.  We will  recall  this  equivalence  below.  Such a surface  can be
obtained  by blowing up 9 points in ${\bf  P}^2$.  (Blowing  down the  section
yields a del Pezzo  surface  $dP_8$, the blow up of ${\bf  P}^2$ at 8 points.)
It can also be given in  Weierstrass  form  $X':  \ \  y^2=x^3+xf'+g'$,  where
$f',g'$ are sections of ${\cal  O}(4),  {\cal  O}(6)$ on the base ${\bf P}^1$.
The recipe of [\ref{MV}] thus extends  naturally:  given $X$ we recover $Z$ as
above,  while the two $E_8$  bundles  are those  which  correspond  to the two
$dP_9$'s:
$$X':  \ \ y^2=x^3+xf'+g', \ \ \ \ \ X^{''}:  \ \ y^2=x^3+xf^{''}+g^{''},$$ 
where 
$$f'=\sum_{i=0}^4 f_it^i, \\ \ g'=\sum_{j=0}^{6} f_jt^j,$$ 
and 
$$f^{''}=\sum_{i=4}^8 f_it^i, \ \ \ g^{''}=\sum_{j=6}^{12} f_jt^j,$$ 
which has the same general form, as seen by exchanging $t$ and $t^{-1}$.

Appealing as it is, this  description  is valid only in a certain  limit.  The
problem is with the  coordinate  $t$ on ${\bf P}^1$:  it is defined only up to
${\bf  P}GL(2,  {\bf C})$.  Actually,  slightly  less is  required:  the three
parameters in $t$ can be identified  as the 2 points $t=0,  t=\infty$,  plus a
scale.  Only the two points are  required  in order to  specify  the upper and
lower parts  $X',X""$,  while the rescaling acts on each by  isomorphisms.  So
let  $\tl{\cal  M}_F$  denote the ${\bf P}^1$  bundle over ${\cal  M}_F$ whose
fiber  over $X \in {\cal  M}_F$ is the  ${\bf  P}^1$  which is the base of the
elliptic  fibration on $X$.  Then  $\tl{\cal  M}_F \x _{{\cal  M}_F}  \tl{\cal
M}_F$ is a ${\bf P}^1 \x {\bf P}^1$ bundle over ${\cal M}_F$, and the space we
need is the open subset  $\tl{\tl{\cal  M}}_F \subset \tl{\cal M}_F \x _{{\cal
M}_F} \tl{\cal  M}_F$  obtained by removing the diagonal in ${\bf P}^1 \x {\bf
P}^1$.  The above  Morrison-Vafa-Friedman-Morgan-Witten  recipe gives a global
map  $\tl{\tl{\cal  M}}_F \ra {\cal  M}_H$.  In order to map  ${\cal  M}_F$ to
${\cal M}_H$ we need a section ${\cal M}_F \ra \tl{\tl{\cal M}}_F$.  This does
{\em not}  exist  globally,  but it can be given on a  codimension-1  boundary
stratum.  In ${\cal M}_F$ this stratum  corresponds to degenerations of $X$ to
the union $X' \cup  X^{''}$ of two  $dP_9$'s  intersecting  along an  elliptic
fiber.  the picture is that the base ${\bf P}^1$ degenerates into the union of
two ${\bf P}^1$'s  intersecting  in a point, the bundles  ${\cal  O}(8), {\cal
O}(12)$ go into $({\cal O}(4), {\cal O}(4))$ and $({\cal O}(6), {\cal O}(6))$,
and $X$ goes to an elliptic  fibration over this singular base.  In this limit
both  the  points   $t=0,   \infty$  go  towards  the   singular   point.  The
corresponding  locus in ${\cal M}_H$ is where the size of the  elliptic  curve
$Z$ goes to $\infty$.  This gets rid of the only non-algebraic parameter, thus
allowing for an algebraic interpretation of the duality.  Explicitly, consider
a two-parameter family of $K3$ surfaces $X_{a,b}$ where $f_i$ is multiplied by
$a^{4-i}$  for $i \leq 4$ and by $b^{i-4}$ for $i \geq 4$, and  similarly  for
the $g_j$.  Then, because of the  above-mentioned  scale parameter,  $X_{a,b}$
depends up to  isomorphism  only on the  product  $ab$.  The  degeneration  is
obtained by letting $a \ra 0$ with $b=1$, or equivalently by letting $b \ra 0$
with  $a=1$.  From the  point  of view of  either  of the two  limiting  ${\bf
P}^1$'s,  ``half the $K3$" (e.g.  12 of the 24 singular  fibers)  has moved to
$\infty$.

\section{Heterotic moduli in six dimensional compactifications}

In six  dimensional  compactifications,  we see  integrable  systems  come  up
naturally on both sides.  We will explore these integrable systems in the next
two sections.

On the  heterotic  side, the moduli  space is  fibered  over the  moduli of K3
surfaces  $Z$.  The fiber over a given  surface $Z$ is the moduli space $\mgz$
of  semistable  $G$-bundles  on $Z$.  Now it was shown by Mukai  [\ref{Mukai}]
that $\mgz$ is a holomorphic  symplectic  manifold,  and in fact it is an {\em
algebraically  integrable  system.}  This  means  that  it  admits  a  natural
fibration $p:  \mgz \ra S$ whose fibers are  Lagrangian  (=maximal  isotropic)
subvarieties  $P_s:=p^{-1}(s)$,  and for  generic  $s \in S$,  each  connected
component  of the fiber $P_s$ is an abelian  variety.  We will  describe  this
structure in a way which extends to other dimensions.  The heterotic fibration
$p$ sends a  $G$-bundle  to {\em the family of its  restrictions  to  elliptic
fibers of $\pi$}.  To make sense of this, consider for each elliptic curve $E$
the moduli space $\mge$ of semistable $G$-bundles on $E$.  Given the family of
elliptic  curves  $\pi:  Z \ra B$, we can put  the  family  of  moduli  spaces
together  to obtain a  fibration  $\mgzb \ra B$ whose  fiber over $b \in B$ is
$\mgeb$.  Now the {\em base} of the heterotic fibration $p$ is
$$S := \{  {\rm {sections:} } \ B \ra \mgzb \} .$$

In  [\ref{FMW}]  it is  observed  that this base $S$ is a weighted  projective
space.  This follows immediately from the fact that, by a theorem of Looijenga
[\ref{L}],  $\mge$ is itself a weighted  projective  space.  There are several
geometric  interpretations for a point $s \in S$; depending on your taste, you
can think of it as  parametrizing a cameral cover  $\tl{\pi}:\tl{B}  \ra B$, a
spectral cover  $\ov{\pi}:\ov{B} \ra B$, or a del Pezzo fibration $\pi:  U \ra
B$.

Cameral covers were  introduced in [\ref{DD}],  and were related to Higgs-type
integrable  systems  in  [\ref{D2}]  and  to  principal  bundles  on  elliptic
fibrations in [\ref{D}].  A cameral  cover is a Galois cover  $\tl{\pi}:\tl{B}
\ra B$ with group $W$, the Weyl group of $G$.  There are some  restrictions on
the local  structure:  it is  locally  pulled  back  from a  certain  standard
$W$-cover.  One way to describe  this  standard  cover is as the map $\mte \ra
\mge$, where $\mte$ is the moduli  space of  $T$-bundles  on $E$, with $T$ the
maximal  torus of $G$.  Specifying a  $T$-bundle,  on any variety  $Y$, is the
same as giving a homomorphism  $\Lambda \ra Pic(Y)$,  where  $\Lambda  \approx
Hom(T,  {\bf  C}^*)$  is the  lattice  of {\em  characters}  of $G$.  So $\mte
=Hom(\Lambda,  E)$ and two points  here map to the same point in $\mge$ if and
only if they  differ  by the  action  of $W$ on  $\Lambda$.  A point $s \in S$
determines both a cameral cover  $\tl{\pi}:\tl{B} \ra B$ and a $W$-equivariant
family of maps  $v_{\lambda}  :  \widetilde{B}  \ra Z$, depending  linearly on
$\lambda \in \Lambda$ and commuting with the projection to $B$.  In fact, this
data suffices to determine $s \in S$ uniquely.  (Actually, some care is needed
in order to  handle  the  possibility  of a  semistable  bundle  on $Z$  whose
restrictions to some, or even all, the elliptic fibers may be unstable.)

Spectral  covers are  quotients  of the cameral  cover by the action of a Weyl
subgroup  $W_0  \subset  W$.  They  can  be  identified   with  the  image  of
$v_{\lambda}$  for any ${\lambda}$  which is fixed  precisely by $W_0$.  It is
often easier to write down explicit  families of spectral  covers, for example
as linear  systems of divisors in $Z$.  On the other hand, the cameral  covers
are somewhat  more  natural,  and  consequently  are easier to use to obtain a
precise  desccription of the fibers.  We will discuss del Pezzo fibrations and
see some concrete examples of spectral covers below.

To complete our  description  of the heterotic  integrable  system, we need to
identify the fibers of the integrable  system $p:  \mgz \ra S$.  This was done
in [\ref{D}], based on [\ref{D2}]:  the fiber  $p^{-1}(s)$  corresponding to a
cameral  cover  $\tl{\pi}:\tl{B}  \ra B$ is given  by the  {\em  distinguished
Prym}:
$$ Prym_{\Lambda}(\tl{B}') := Hom_W(\Lambda, Pic {\tl B}).$$
When $ {\tl B}$ is a non-singular  projective variety (a curve, in our current
dimension),  $Prym_{\Lambda}(\tl{B}')$  will be the  product of its  connected
component $Prym^0_{\Lambda}(\tl{B}')$, an abelian variety which up to a finite
cover is an abelian  subvariety of  $Pic^0(\tl{B})$,  with a discrete  abelian
group of  connected  components.  (There is a fine point  here,  which is that
$p^{-1}(s)$   is   naturally    identified    with   a   {\em   torser}   over
$Prym_{\Lambda}(\tl{B}')$;  this means that (if nonempty) they are isomorphic,
but not  naturally;  so the  family of all  $Prym_{\Lambda}(\tl{B}')$  will be
different than the total space $\mgz$.)

\section{F-theoretic moduli in six-dimensional compactifications}

The continuous moduli which occur in six-dimensional compactifications of 
F-theory include the complex structure of the Calabi-Yau threefold $X$, the 
non-zero holomorphic volume form $\omega_X \in H^{3,0} = H^0(K_X)$, and 
the Ramond-Ramond fields which specify a point of $H^3(X, {\bf R})$ modulo 
$H^3(X, {\bf Z})$. This torus, called the intermediate Jacobian of $X$,  
has a natural complex structure,
$$J^3(X) := H^3(X, {\bf Z }) \backslash H^3(X, {\bf C }) / 
(H^{30} \oplus H^{21}).$$
According to [\ref{BB}],[\ref{W4fl}], the full moduli space for 
F-theory compactifications to 4 dimensions includes some discrete 
parameters as well; the intermediate Jacobian should be replaced by the 
{\em Deligne cohomology group} :
$$ 0 \ra J^3(X) \ra {\cal D} \ra H^{2,2}(X,{\bf Z}) \ra 0, $$
or rather by the part of the Deligne cohomology mapping to the 
primitive cohomology $H^{2,2}_0(X,{\bf Z})$. In six dimensional 
compactifications we may expect this primitive group, in generic situations, 
to be finite; nevertheless, it is natural to include it in our moduli.

It was  proved in  [\ref{DM}]  that this full  moduli  space -  including  the
complex  structure  parameters, the  holomorphic  volume form, and the Deligne
cohomology  - is a  holomorphic  integrable  system.  The origin and nature of
this integrable system seem very different than those of the heterotic system.
One obvious  distinction is that the heterotic  fibers are abelian  varieties,
that is they come with definite  polarizations; but the intermediate Jacobians
obtained   on  the   F-theory   side  are  non   algebraic   varieties:  their
polarizations are indefinite, of signature $(1,h^{21})$.

We sketch  one way to see that  this is indeed an  integrable  system.  Let us
start  with a very  general  question:  given  a  family  ${\cal  X} \ra S$ of
$g$-dimensional,  polarized complex tori over a  $g$-complex-dimensional  base
$S$, is there a holomorphic symplectic form $\Sigma$ on $\cal X$ such that the
fibers are Lagrangian?

The answer is that such a $\Sigma$  corresponds to a field of symmetric  cubic
tensors on the base $S$ satisfying an  integrability  condition  which ensures
that it can be given  locally  as third  partials  of a single  function,  the
prepotential.

Indeed, given $\Sigma$ we get an isomorphism of the holomorphic tangent bundle
$T(S)$ of the base with the relative  cotangent bundle $T^*({\cal X}/S)$ along
the fibers $X_s, s \in S$.  By partitioning a basis of $H^1(X_s,{\bf Z})$ into
``$\alpha$" and ``$\beta$" isotropic subspaces (for some $s_0 \in S$ and hence
via the  Gauss-Manin  connection  for  all  nearby  $s \in  S$) we get a local
holomorphic  trivialization  of $T^*({\cal  X}/S)$, and hence of $T(S)$.  this
allows us to  identify  $S$ locally as an open  subset of a vector  space $V$.
The period matrix of the fiber $X_s$ then lives  naturally in $Sym^2V^*$.  The
holomorphic  structure of the fibration  ${\cal X} \ra S$ together with such a
trivialization is given by its period map
$$p : S \ra Sym^2V^*.$$
At each  point  of $S$,  the  differential  of the  period  map, $ dp :  V \ra
Sym^2V^*$,  can be  identified  as an element of $V^* \otimes  Sym^2V^*$.  The
result proved in [\ref{DM}]  is that this ${\cal X}$ admits a symplectic  form
$\Sigma$  inducing the given  trivialization  and for which the  fibration  is
Lagrangian if and only if $dp$ is in the subspace  $Sym^3V^*$  of $V^* \otimes
Sym^2V^*$.  In this case there exists locally a  prepotential  function  whose
second  partials give the period matrix $p$, so the third  partials give $dp$,
the cubic.

In our case, let $\cal M$ be the moduli space of complex structures 
on $X$, and let
$S := \tl{\cal M}$ be the moduli space of complex structures 
on $X$ together with a nonzero
holomorphic volume form $\omega$, so $S$ is a $C^*$-bundle 
over $\cal M$. The cubic is, of course,
Yukawa's:
$$Sym^3T_s \tl{\cal M} \ra Sym^3T_s {\cal M} = Sym^3H^1(X_s, TX_s) \ra 
H^3(X_s, \wedge ^3TX_s) \ra H^3(X_s, {\cal O}_{X_s}) \ra {\bf C} $$ 

where each of the last two maps is given by  contraction  with the volume form
$\omega$.  This then gives rise to the  holomorphically  integrable  system of
intermediate  Jacobians.  For  extension  to the family of Deligne  cohomology
groups, an additional fact is needed, namely that the Abel-Jacobi image of any
family of  null-homologous  algebraic  cycles is an  isotropic  subspace  with
respect  to the  symplectic  structure  just  constructed  on  the  family  of
intermediate Jacobians.

\section{Duality correspondence along the base}

As  explained  earlier,  a point  in the  base  of  either  integrable  system
corresponds  to a  family,  parametrized  by the  heterotic  base  $B$, of the
eight-dimensional data.  The correspondence along the base is thus obtained by
putting the  8-dimensional  correspondence  into families.  On the F side, the
base is just the moduli space of Calabi-Yaus (with a holomorphic volume form).
On the heterotic side there are three useful  descriptions  for a point of the
base:  it represents  either a cameral cover (with the additional  data of the
evaluation  map $v:  \Lambda \x \tl{B} \ra Z$), or a spectral  cover, or a del
Pezzo fibration.

We have already  discussed  cameral and spectral  covers.  When the  structure
group is of type $A_{n-1}$, the cameral cover is an  $n!$-sheeted  cover.  The
basic spectral cover $\ov{B}$,  corresponding to the first fundamental weight,
is   $n$-sheeted,   and  the  one   corresponding   to  the  k-th   weight  is
$(^n_k)$-sheeted  (one of its  points is given by an  unordered  $k$-tuple  of
points in a fiber of  $\ov{B}$).  For  $E_8$,  the  cameral  cover has  degree
${\#}W = 696,729,600=  2^{14}3^55^27$,  while the smallest  spectral cover has
degree 240.

Let $D$ be a del Pezzo surface $dP_8$.  The blowup picture identifies  $H^2(D,
{\bf Z})$ with the  indefinite  lattice ${\bf  Z}^{(8,1)}$,  so the  primitive
cohomology  $H^2_0(D, {\bf Z})$ is identified with the $E_8$ lattice  $\Lambda
\subset {\bf  Z}^{(8,1)}$;  this  identification  is determined only up to the
action of $W$.  An  embedding  of an  elliptic  curve $E$ as an  anticanonical
divisor in $D$ then induces a  homomorphism  $\Lambda \ra Pic^0(E) = E$, up to
the $W$ action on $\Lambda$.  But as we saw, a  homomorphism  gives a point of
$Hom(\Lambda,  E) = \mte$, and two points here map to the same point in $\mge$
if and  only  if  they  differ  by the  action  of $W$ on  $\Lambda$.  So  the
embedding of $E$ in $D$ is exactly equivalent to specifying a point of $\mge$.
Explicitly,  an $E_8$  bundle on $E$ is given by 8 line  bundles  of degree 0,
hence 8 points of $E$; the  embedding  of $E$ into $D$ is  recovered  by first
embedding  $E$ into ${\bf P}^2$ by the linear  system $3o$, where $o \in E$ is
the  origin, and then  blowing up ${\bf P}^2$ at the images of the 8 points of
$E$.  This  construction  works  well  in  families   [\ref{CD}],  giving  the
interpretation  of a point of the  base  $S$ as a  fibration  $U \ra B$  whose
fibers  are  $dP_8$  surfaces  together  with  a  (relatively   anticanonical)
embedding of the heterotic  Calabi-Yau $Z$ into $U$.  Given this  description,
we recover the spectral cover as parametrizing the 240 lines in each del Pezzo
fiber, while the cameral cover parametrizes 8-tuples of disjoint lines.

In case the  structure  group of the bundle is the full $E_8$, the generic del
Pezzo fiber will be  non-singular.  A  singularity  of type ADE on the generic
del Pezzo implies that the structure group of the equivalent  cameral cover is
reduced  to the Weyl  group of the  complementary  subgroup  in $E_8 \x  E_8$.
(This can also be expressed in terms of the  spectral  cover,  which will then
have the zero  section as a component  with  appropriate  multiplicity.)  This
results in a reduction of the structure  group of the heterotic  bundle to the
complementary  subgroup in $E_8 \x E_8$.  This reduction  shows up as enhanced
symmetry of the compactified theory, so it must be visible also in a family of
ADE singularities of the corresponding type in the F-theoretic Calabi-Yau.  We
conclude that the two $n$-dimensional objects:  the F-theoretic Calabi-Yau $X$
and the heterotic  pair-of-delPezzo-fibrations  $U' \cup U^{''}$ must have the
{\em same} types of  singularities.  (Except that the heterotic object is also
singular  along the divisor $Z$.)  In the  geometric  limit, this  matching is
evident, as the two varieties are actually  isomorphic  here.  As we move into
the interior of moduli  space by  returning  to finite  volume, the  heterotic
object does not  change.  The  singularity  of the  F-theoretic  $X$ along $Z$
disappears, but the other singularities are supposed to remain.  Much evidence
for this is presented in [\ref{BIKMSV}], [\ref{AG}], and [\ref{FMW}].

\section{Duality correspondence along the fibers}

Does  duality  preserve  the  integrable  system  structures?  Not quite.  One
obvious  difference is in the dimensions:  the F-theoretic  integrable  system
involves all the moduli, while on the heterotic  side the system itself is the
fiber of a map of the  heterotic  moduli  to the  moduli  of  $K3$.  A  second
important  difference  was  already  noted  above:  the  heterotic  fibers are
algebraic (abelian) varieties, but the intermediate  Jacobians on the F-theory
side are non  algebraic  complex  tori, with  ``Lorentzian"  polarizations  of
signature $(1,h^{21})$.

Going to the ``geometric" limit which was described in the eight-dimensional 
context resolves both of these difficulties. On the heterotic side, 
we take the Kahler metric on $B$ as well as the size of the elliptic 
fibers to be large. On the F side The Calabi-Yau $X$ splits into 
$X' \cup X^{''}$ which is an elliptic fibration over $B_F^{'} \cup B_F^{''}$, 
while the latter becomes a ${\bf P}^1 \cup {\bf P}^1$ bundle over $B$. 
The intersection $X' \cap X^{''}$ is again the heterotic $Z$. This 
degeneration kills the $H^{30}$ term, so the surviving part of the 
intermediate Jacobian,
$$ J^3(X) \ra J^3(X') \x J^3(X^{''})$$
is now indeed an abelian variety. The result, proved in [\ref{CD}], is
that the continuous part $ J^3(X') \x J^3(X^{''})$ of the F-theory moduli 
is identified with the continuous part 
$Prym_{\Lambda}^0(\tl{B}') \x  Prym_{\Lambda}^0(\tl{B}^{''})$
of the heterotic moduli, while the F-theory discrete part embeds as a 
subgroup of finite index in the corresponding heterotic discrete group.
This possible finite index may well vanish in general, yielding an actual 
isomorphism; in our approach this depends on the vanishing of certain 
(finite) group cohomologies which have
not yet been computed. In the physically important case that the base $B$ is 
${\bf P}^1$, such an isomorphism was obtained in [\ref{K}] for structure groups
$E_n,~~n \leq 7$. Another result over ${\bf P}^1$, but valid also for $E_8$, 
was recently announced in [\ref{FMW2}].

The result proved in [\ref{CD}] can actually be stated {\em without} 
going to the limit: it identifies the Deligne group ${\cal D}(U')$ of 
the heterotic del Pezzo fibration as a subgroup of finite index in 
the heterotic fiber moduli, which are given by 
$Prym_{\Lambda}^0(\tl{B}')$. The additional feature present in the limit 
is that the heterotic and F-theoretic del Pezzo fibrations can be 
identified, so the general algebro-geometric result becomes relevant to 
the matching of the fibers between the two theories.

\section{Four dimensional compactifications}

The analysis of cameral and spectral covers and del Pezzo fibrations  works in
any dimension.  It allows  parametrization of bundles on $Z$ whose restriction
to each elliptic  fiber is  semistable,  together  with a bit of extra data, a
regularization [\ref{D}].  In particular, the moduli space of these bundles is
still  fibered  over the space of covers, and the fiber is still  given by the
distinguished  Prym,  $Prym_{\Lambda}(\tl{B})$.  The result of  [\ref{CD}]  is
also still valid:  the Deligne group of the del pezzo  fibration a subgroup of
finite index in $Prym_{\Lambda}(\tl{B})$.  For a bundle whose restriction to a
particular elliptic fiber $E$ is not semistable, the corresponding spectral or
cameral  "cover"  will  contain  the entire  fiber  $E$, so it is no longer an
(everywhere  finite)  cover.  It seems that the  spectral  description  of the
heterotic  moduli can be extended, with  modifications,  to include this case,
but further work needs to be done to  substantiate  this.  Various  results on
four dimensional  compactifications  can be found in  [\ref{FMW}],[\ref{BJPS}]
and elsewhere.

\noindent{\bf Acknowledgements}

It is a pleasure to thank Paul Aspinwall,  Pierre  Deligne,  Victor  Ginzburg,
Antonella  Grassi,  Dave  Morrison,  Tony Pantev and Edward  Witten for useful
discussions.

\bigskip
\bigskip

\section*{References}
\begin{enumerate}

\item
\label{FMW}
R. Friedman, J. Morgan and E. Witten, {\it Vector Bundles and 
F-Theory}, Commun. Math. Phys. {\bf 187} (1997) 679, hep-th/9701162.

\item
\label{D}
R.Y. Donagi, {\it Principal bundles on elliptic fibrations}, 
Asian J. Math. {\bf1} (1997), 214-223, alg-geom/9702002.

\item
\label{BJPS}
M. Bershadsky, A. Johansen, T. Pantev and V. Sadov, {\it On 
four-dimensional Compactifications of F-theory}, hep-th/9701165.

\item
\label{DD}
R.Y. Donagi, {\it Decomposition of Spectral covers}, 
Journees de Geometrie Algebrique d'Orsay, Asterisque 218
(1993) 145.

\item
\label{D2}
R.Y. Donagi, {\it  Spectral covers, in: Current topics in complex algebraic 
geometry}, MSRI pub. {\bf 28} (1992), 65-86, alg-geom 9505009.

\item
\label{V}
C. Vafa, {\it Evidence for $F$-theory}, Nucl. Phys. {\bf B 469} (1996) 
403, hep-th/9602022.

\item
\label{MV}
D. Morrison and C. Vafa, {\it Compactifications of F-theory on Calabi-Yau
Threefolds I, II}, Nucl. Phys. {\bf B 473} (1996) 74; ibid. {\bf B 476}
(1996) 437.

\item
\label{BIKMSV}
M. Bershadsky, K. Intrilliator, S. Kachru, D.R. Morrison, V. Sadov and
C. Vafa, {\it Geometric Singularities and Enhanced Gauge Symmetries},
hep-th/9605200.

\item
\label{AG}
P. Aspinwall and M. Gross, {\it The $SO(32)$ Heterotic String on a K3
surface}, hep-th/9605131.

\item
\label{CD}
G. Curio and R.Y. Donagi, {\it Moduli in heterotic/F-theory duality}, 
hep-th/9801057.

\item
\label{N}
K.S. Narain, {\it New heterotic string theories in uncompactified dimensions 
less than 10}, Phys. Let. 169B(1986), 41-46.

\item
\label{Mukai}
S. Mukai, {\it Symplectic structure of the moduli space of sheaves on an 
abelian or K3 surface}, Inv. Math. 77(1984), 101-116.

\item
\label{L}
E. Looijenga, {\it Root systems and elliptic curves}, Inv. Math. 38 
(1976), 17-32, and {\it Invariant theory for generalized root systems}, 
Inv. Math. 61 (1980), 1-32.

\item
\label{BB}
K. Becker and M. Becker, {\it {\cal M}-theory on eight-manifolds}, 
hep-th/9605053.

\item
\label{W4fl}
E. Witten, {\it On Flux Quantization in M Theory and the Effective 
Action},
hep-th/9609122.

\item
\label{DM}
R. Donagi and E. Markman, {\it Cubics, Integrable Systems, and Calabi-Yau 
Threefolds},in: Proceedings of the Hirzebruch 65 Conf. on Alg. Geom, 1993, 
ed. M. Teicher, Israel Math. Conf. Proc. 9, alg-geom 9408004.

\item
\label{K}
V. Kanev, {\it Intermediate Jacobians and Chow groups of threefolds with a 
pencil of del Pezzo surfaces}, Annali di Matematica pura ed applicata (IV), 
Vol. CLIV (1989) 13.

\item
\label{FMW2}
R. Friedman, J. Morgan and E. Witten, {\it Principal G-Bundles over
elliptic curves}, alg-geom/9707004.

\end{enumerate}
\end{document}